\documentclass{taiwan}
\input{psfig}

\begin{document}

\title{$D^{\circ}-{\bar D^{\circ}}$ Mixing and Rare Decays}

\author{Hwanbae Park}

\address{Department of Physics, Korea University, Seoul, 
Korea\\E-mail: hbpark@hep.korea.ac.kr}


\maketitle\abstracts{ 
Current status of charm mixing search, lifetime difference measurement and
rare decay search are reported.
The best upper limit at $95\%$ CL for mixing is reported from the CLEO
collaboration.
E791 has reported lifetime difference measurement 
and results of rare decay searches. 
Rare decay searches of the FOCUS collaboration  are in progress  
and their sensitivities to branching ratios are at the level of a few $\times 10^{-6}$.}

The Standard Model (SM) predictions of charm mixing and rare decays are negligibly small
and new physics effects are not hidden by the SM effects.
There are new physics scenarios such as a large mass of fourth down-quark,
two higgs-doublet model which gives an additional box diagram by replacing 
${\rm W}^{\pm}$ with the charged Higgs, flavor changing neutral Higgs
(FCNC) model in which up-quark sector is treated differently from down-type sector.
There are also
leptoquark model in which scalar leptoquark bosons participate in a mass
difference  via exchange inside a box diagram, and supersymmetry model which gives
an addtional box diagram contribution.~\cite{NEWPHY}
These scenarios predict large mixing effects and
any experimental observations in searches of mixing and rare decay would  
indicate the existence of new physics beyond the Standard Model.

\section{Mixing}
Mixing occurs because two weak eigenstates are not the mass eigenstates.
The probability of finding $\bar D^{\circ}$ at time t with produced $D^{\circ}$ at $t=0$ 
is
\begin{equation}
 \Gamma (D^{\circ}\rightarrow \bar D^{\circ})
   = {1\over 4}{|{q\over p}|}^2e^{-\Gamma_+t}
   [1 + e^{\Delta\Gamma t}
                 -2e^{({1\over 2}\Delta\Gamma t)}{\cos(\Delta m t)}] 
\label{eq:mixprob}
\end{equation}
\noindent
in which two states oscillate with rate expressed by $\Delta m$ and $\Delta\Gamma$.
\noindent
$\Delta m\ll\Gamma$ and $\Delta\Gamma\ll\Gamma$ are good approximations for the charm
particle and 
the mixing rate is  
\begin{equation}
r_{\rm mix} = {\Gamma(D^{\circ}\rightarrow {\bar D^{\circ}}
                     \rightarrow {\bar f})\over {
                     \Gamma(D^{\circ}\rightarrow f)}} =
                    {1\over 4}e^{-\Gamma t}{|{q\over p}|}^2(x^2+y^2)\Gamma^2t^2
\label{eq:mixrate}
\end{equation}
\noindent
with following definition of the mixing parameters,
\begin{equation}
 x={\Delta m\over{\Gamma}};~~~ 
y={\Delta\Gamma\over{2\Gamma}}
\label{eq:mixpara}
\end{equation}
\noindent
\noindent
\noindent
$\Delta\Gamma$ is expected to be of the same order of magnitude as  $\Delta m$, 
and new physics or long distance contribution effects
are sensitive to $\Delta m$.

Mixing is studied by searching for wrong sign signals or measurement of 
lifetime difference in CP eigenstates.
The direct search is to look for the produced $D^{\circ} (\bar D^{\circ})$ 
that decays as $\bar D^{\circ} (D^{\circ})$.
The flavor of the produced $D$ is tagged by the charge of the bachelor pion
in $D^{\star +}\rightarrow D^{\circ}\pi^+$, and $D^{\circ}$ then decays hadronically 
$(K^+\pi^-,~K^+\pi^-\pi^+\pi^-)$ and semileptonically $(K^+\ell^-\bar{\nu_{\ell}})$.
\noindent
Since the Doubly Cabbibo Suppressed Decay (DCSD) gives the wrong sign signals
as well as mixing, and the charm mixing in the SM is expected to be small,
the DCSD contribution in the hadronic modes may not be neglected.  
The amplitude of the mixing is given by
\begin{equation}
A_{\rm WS}=A_{\rm DCSD}(D^{\circ}\rightarrow\bar f)+
            A_{\rm mix}(D^{\circ}\rightarrow{ \bar D^{\circ}}\rightarrow\bar f )
\label{eq:mixpol}
\end{equation}
\noindent
The wrong sign decay rate is then
\begin{equation}
\begin{array}{rcl}
r_{\rm WS} & = &
              {1\over 4}e^{-\Gamma t}  <{\bar f}|H|{\bar D^{\circ}}>_{\rm CF}^2 
              {|{q\over p}|}^2\times  \\
    & & ~~~~~[4{|\lambda|^2}  +
    (2{\it Re}(\lambda)\Delta\Gamma +
               4{\it Im}(\lambda)\Delta m)t +  
   (\Delta m^2 + {1\over 4}\Delta\Gamma^2)t^2 ]\\ & &
\end{array}\label{eq:mixpoleq}
\end{equation}
\noindent
 where
\begin{equation}
 \lambda ={p\over q}{ <{\bar f}|H|D^{\circ}>_{\rm DCSD}\over
                         {
                           <{\bar f}|H|{\bar D^{\circ}}>_{\rm CF}}
                         }
\label{eq:lambda}
\end{equation}
\noindent
The significant differences in the structure of the proper decay time
for the DCSD, the interference between DCSD and mixing,
and the mixing allow us to estimate their respective contributions.

\noindent
The CP conjugate rate is
\begin{equation}
\begin{array}{rcl}
{\bar r}_{\rm WS} & = &
              {1\over 4}e^{-\Gamma t}<{ f}|H|{{D^{\circ}}}>_{\rm CF}^2
              {|{p\over q}|}^2\times  \\
    & & ~~~~~[4{|{\bar \lambda}|^2} +
    (2{\it Re}({\bar\lambda})\Delta\Gamma +
               4{\it Im}({\bar\lambda})\Delta m)t +
                (\Delta m^2 + {1\over 4}\Delta\Gamma^2)t^2 ]\\ & &
\end{array}\label{eq:mixpoleq1}
\end{equation}
\noindent
where
\begin{equation}
                {\bar \lambda} ={q\over p}{ < f|H|{\bar D^{\circ}}>_{\rm DCSD}\over
                        {
                         < f|H|{D^{\circ}}>_{\rm CF}}
                        }
\label{eq:cpcon}
\end{equation}
\noindent
With definition of the CP violation angle $\phi$ and strong phase $\delta$:
\begin{equation}
e^{-\it i\phi}={p\over q};~~~
\sqrt{R_{\rm DCSD}}\cdot e^{-\it i\delta}=
    {{<\bar f|H|D^{\circ}>_{\rm DCSD}}\over
     {<\bar f|H|{\bar D^{\circ}}>_{\rm CF}}}
\label{eq:phase}
\end{equation}

\noindent
Eq.~(\ref{eq:mixpoleq}) is simplied
\begin{equation}
\begin{array}{rcl}
 r_{\it WS}& = & e^{-\Gamma t}<\bar f|H|{ \bar D^{\circ}}>_{\it CF}^2\times \\
          &&~~~~ [R_{\it DCSD}  \\ 
   &&~~~~+(y\sqrt{R_{\it DCSD}}\cdot\cos(\delta\pm\phi) - x\sqrt{R_{\it DCSD}}\cdot\sin
    (\delta\pm\phi))\Gamma t  \\
   &&~~~~ + {1\over 2}R_{\rm mix}(\Gamma t)^2
  ]
\end{array}\label{eq:mixf}
\end{equation}      

\noindent
Defining $x^{\prime}= x\cdot\cos\delta + y\cdot\sin\delta$ and 
$y^{\prime}= y\cdot\cos\delta - x\cdot\sin\delta$, and assuming CP invariance ($\phi=0$),
Eq.~(\ref{eq:mixf}) then becomes
\begin{equation}
r_{\rm WS}\propto e^{-\Gamma t}
      [R_{\rm DCSD}+y^{\prime}\sqrt{R_{\rm DCSD}}(\Gamma t)
                   +{1\over 2}R_{\rm mix}(\Gamma t)^2]
\label{eq:xyparam}
\end{equation}
\noindent
The dependence on  $x^{\prime}$ and $y^{\prime}$ is distingushable
due to the interference with direct decay amplitude, which induce the linear
interference.

\section{Mixing Searches}
\subsection{Wrong Sign Searches}
The ALEPH has used $4\times 10^6$ hadronic $Z^{\circ}$ events collected
from 1991 to 1995 and searched the wrong sign signals in
$D^{\star +}\rightarrow D^{\circ}\pi^+$ with $D^{\circ}$ 
decaying to $K^+\pi^-$.~\cite{ALEPH}
\begin{figure}[t]
\psfig{figure=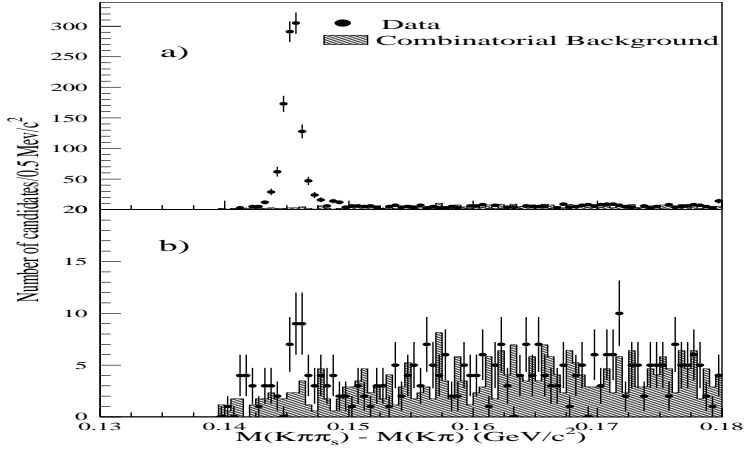,height=1.2in,width=5.5in}
\vskip 2.5cm
\caption{ALEPH mixing search in the $K\pi$ mode. (a) is the right sign
and (b) is the wrong sign distributions.
The hatched histogram is the combinatorial background and  is assumed
same shape from events in the sideband region of $D^{\circ}$ mass 
above $2.1 {\rm GeV}/c^2$,
and is normlized to the number of candidates having $\Delta M > 160 {\rm MeV}/c^2$.
\label{fig:aleph_mix1}}
\end{figure}
The right sign and wrong sign mass plots are shown in Fig.~\ref{fig:aleph_mix1}.
$N_{RS}=1038.8\pm 32.5\pm 4.3$ and $N_{WS}=19.1\pm 6.1\pm 3.5$ 
after combinatorial and physics backgrounds 
substraction are obtained.
They also measured a relative branching ratio of $B(D^{\circ}\rightarrow K^+\pi^-) /
  B(D^{\circ}\rightarrow K^-\pi^+)  =(1.84\pm 0.59\pm 0.34)\%$ and
set $r_{\rm mix} < 0.92\%$ at $95\%$ CL on the mixing rate assuming 
no interference between DCSD and mixing.

The CLEO has used $9.0~{\rm fb}^{-1}$ 
of $e^+e^-$ data 
with CLEO II.V detector from 1996 to 1999 runs.
The analysis takes great advantage of the silicon vertex detector.
\begin{figure}[t]
\vskip 1.5cm
\centerline{\psfig{figure=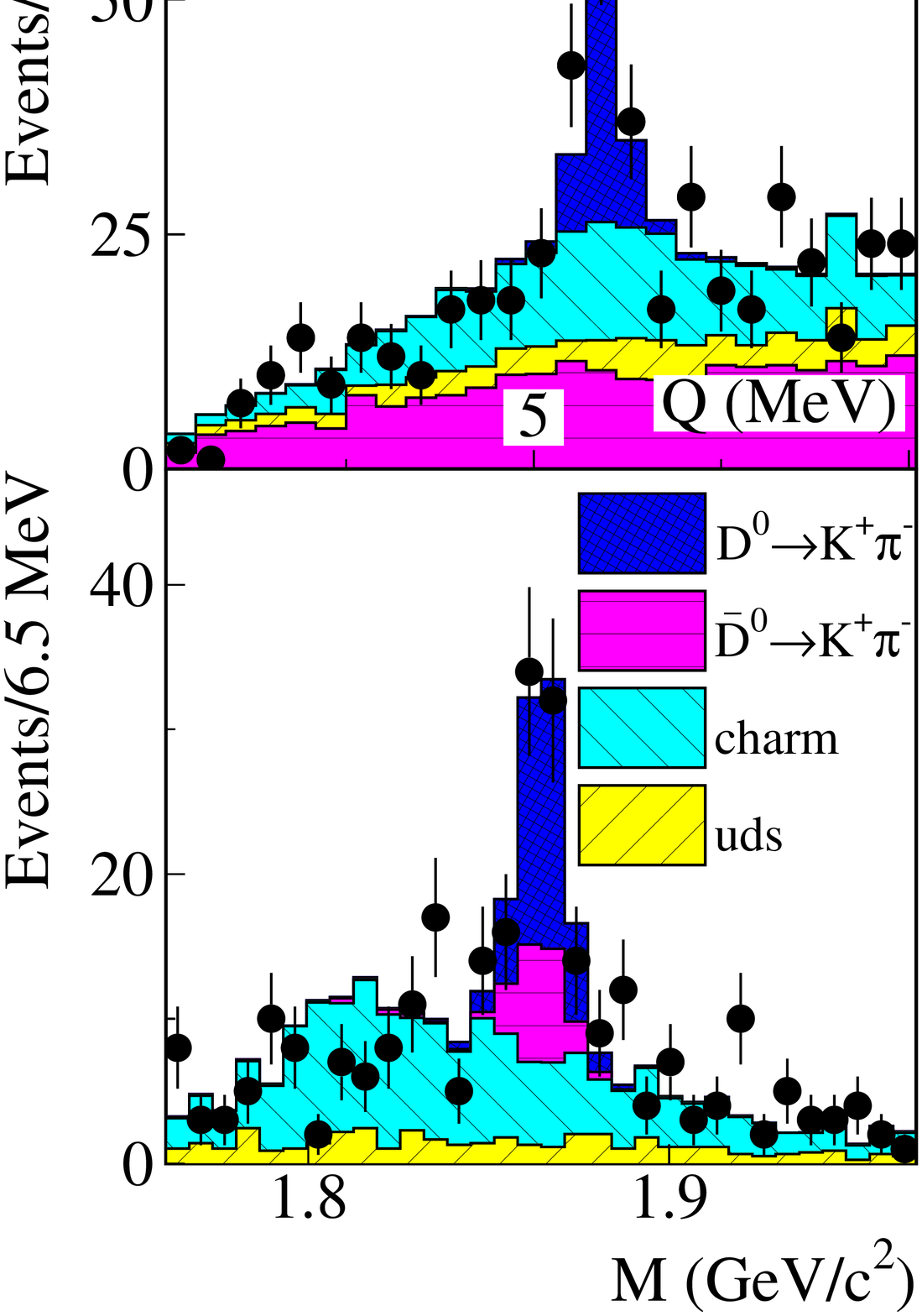,height=1.75in,width=4.5in}}
\caption{CLEO mixing search in $K\pi$ mode. The upper plot is the
mass difference Q ($M_{K\pi\pi}-M_{K\pi}-M{\pi}$) and 
the lower plot is reconstructed $D$ mass distributions of
wrong sign data. Backgrounds from various sources are shown.
\label{fig:prl_col}}
\end{figure}
\noindent
The $D$ mass  and Q ($M_{K\pi\pi}-M_{K\pi}-M{\pi}$) plots for
the wrong sign candidates are shown in Fig.~\ref{fig:prl_col}.
The signal shape is determined by the right sign distribution and the background shapes
are estimated by Monte Carlo simulation. 
$44.8^{+9.7}_{-8.7}$ and $13527\pm 116$ for the wrong sign 
and the right sign events are found from fits, respectively.~\cite{CLEO} 
Using these numbers, they measure
\begin{equation}
{R_{\rm WS}(K\pi)\over{R_{\rm RS}(K\pi)}}=(0.335^{+0.064}_{-0.066}\pm 0.040)\%
\label{eq:CLEOmix}
\end{equation}
\noindent
From fits, they also find one-dimensional intervals at $95\%$ CL
of:
\begin{equation}
{1\over 2}x^{\prime 2}<0.039\%;~~~ -5.6\% < y^{\prime} < 1.3\%
\label{eq:CLEOmixing}
\end{equation}

The E791 is the fixed target hadroproduction experiment at Fermilab which uses
a 500 ${\rm GeV}/c^2~\pi^-$ beam.
They logged $2\times 10^{10}$ hadronic interactions and 
reconstructed more than 200,000 hadronic events for 1990-1991 runs.
The wrong sign signals in hadronic ($K\pi$/$K3\pi$) 
and semileptonic ($K\ell\bar\nu$) decay modes have been searched.
The loose selection criteria are applied to reduce data samples and then
selection cuts are optimized with artificial neural networks.
\noindent
5643 and 3469 of the right sign signals for $K\pi$ and $K3\pi$ 
are obtained, respectively.~\cite{E791HAD}
They first performed most general fit allowing for CP violation, mixing
interference and no model dependence, and set the $90\%$ CL limits:
\begin{equation}
r_{\rm mix}(D^{\circ}\rightarrow{\bar D^{\circ}}) < 1.45\%;~~~
r_{\rm mix}(\bar D^{\circ}\rightarrow D^{\circ}) < 0.74\%
\label{eq:E791mix}
\end{equation}
\noindent
Assuming CP violation only in the interference term, they find
$90\%$ CL limits of:
\begin{equation}
r_{\rm mix} < 0.85\%
\label{eq:mixrest}
\end{equation}
\noindent
They also evaluated relative branching ratio for the DCSD modes with assumption of
no mixing:
\begin{equation}
R_{\rm DCSD}(K\pi)=(0.68^{+0.34}_{-0.33}\pm 0.07)\%;~~~
R_{\rm DCSD}(K3\pi)=(0.25^{+0.36}_{-0.34}\pm 0.03)\%
\label{eq:DCSDmix}
\end{equation}
\noindent
In the semileptonic analysis, $D^{\star +}\rightarrow 
D^{\circ}\pi^+$ with $D^{\circ}$ decaying to
$K^+\ell^- \bar{\nu}_\ell$ is searched.
There is a two-fold momentum ambiguity in the $D^{\circ}$ momentum due to the missing
neutrino and the higher momentum solution is picked from Monte Carlo study.
\noindent
They fit to the mass difference and the decay time distribution with 
$D^{\circ}$ mass fixed.
\begin{figure}[t]
\centerline{\psfig{figure=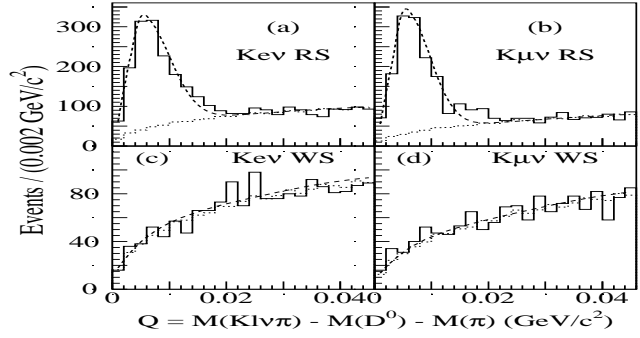,height=1.25in,width=3.5in}}
\vskip-0.5cm
\caption{E791 mixing search with semileptonic modes. The upper
plots are for the right sign and the lower plots are for the wrong
sign data. The dashes are the overplot of fit results and the dots are the 
background shape obtained from combining $D^{\circ}$ from one event 
with bachelor $\pi$ from another event. 
\label{fig:lmix1}}
\end{figure}
\noindent
The right and wrong sign Q plots are shown in Fig.~\ref{fig:lmix1}.
$N_{\rm RS}=1237\pm 45$ for $Ke\nu$ and
$N_{\rm RS}=1267\pm 44$ for $K\mu\nu$ are obtained.~\cite{E791SEMI}
After performing two-dimensional unbinned maximum likelihood fit and correcting for the
different detector acceptance for the different decay time behaviors of the mixed 
and right sign decays,
they obtained:
\begin{equation}
 r_{\rm mix}(Ke\nu)=(0.16^{+0.42}_{-0.37})\%;~~~
 r_{\rm mix}(K\mu\nu)=(0.06^{+0.44}_{-0.40})\%
\label{eq:KMU}
\end{equation}
Using these numbers and they have determined the weighted average and
set $90\%$ CL limit:  
\begin{equation}
r_{\rm mix}=(0.11^{+0.30}_{-0.27})\%;~~~
 r_{\rm mix}<0.5\%
\label{eq:upper}
\end{equation}

The FOCUS is the fixed target photoproduction experiment at Fermilab and is 
the successor to E687.
The data collected during 1996-1997 run have reconstructed 15 times more charm decays 
than the E687.
\begin{figure}[t]
\centerline{\psfig{figure=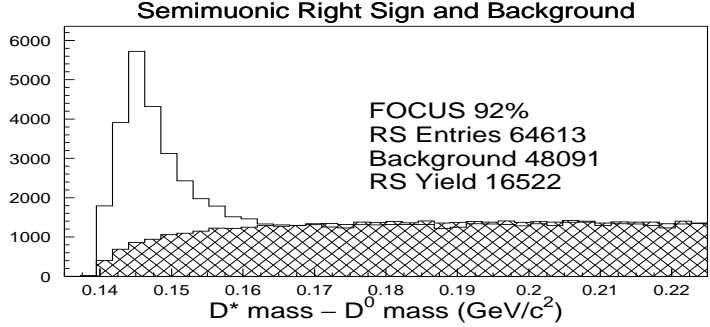,height=1.75in,width=4.5in}}
\caption{FOCUS preliminary mixing search in semimuonic mode.
The hatched background shape is modeled by replacing pions in primary by pions from 
a different event.
\label{fig:mike_ps}}
\end{figure}
\noindent
Data is being used in search of semileptonic and hadronic wrong sign decay modes.
A mass difference plot for their preliminary right sign semimuonic analysis with 
$92\%$ of their total data is shown in Fig.~\ref{fig:mike_ps}.
Approximately 16500 background subtracted right sign candidates are found
and roughly equal numbers of events in the electron mode are obtained as well.
The wrong sign signal region is not looked for until selection cuts have been optimized
and backgrounds understood. 
$r_{\rm mix}$ will be extracted from two dimensional fit to the mass difference and 
the proper time distribution.
At $90\%$ CL limit on a sensitivity extrapolation from the preliminary studies indicate:
\begin{equation}
r_{\rm mix} < 0.12\%
\label{eq:nothing}
\end{equation}
\subsection{Lifetime Difference Searches}
The lifetime difference is obtained by the measurement of the lifetimes 
of CP eigenstates or
comparing number of decays observed at any decay time:
\begin{equation}
\begin{array}{rcl}
& \Gamma(D^{\circ}\rightarrow K^+K^-) & -\Gamma(D^{\circ}\rightarrow K^-\pi^+)
 =  \Gamma_+ - {1\over 2}(\Gamma_+ + \Gamma_-)={1\over 2}\Delta\Gamma \\
& {\rm ln}{{N(D^{\circ}\rightarrow K^+K^-)}\over{N(D^{\circ}\rightarrow K^-\pi^+)}}
 & =  {\rm ln}({A_{KK}\over A_{K\pi}}) - {\Delta\Gamma\over 2}t
\\ & &
\end{array}\label{eq:lifedif}
\end{equation}
\noindent
The E791 measured the lifetimes of the $D^{\circ}\rightarrow K^+K^-$
which is CP even eigenstate and $D^{\circ}\rightarrow K^-\pi^+$ which is CP
mixed state.~\cite{E791Life}
\noindent
After background subtraction $6683\pm 161~K^+K^-$ and
$60571\pm 353~K^-\pi^+$ events are found.
The lifetime is extracted from the exponential fits to measured 
reduced lifetime distribution after particle identification weighting and 
acceptance corrections, and
at $90\%$ CL limit on $y$ is:
\begin{equation}
-0.04 < y < 0.06
\label{eq:whatis}
\end{equation}

The FOCUS will also search for the lifetime difference by comparing
the lifetimes of the CP even final state and CP mixed final state.
\noindent
It is shown that the reflection from $K\pi$ can be controlled.~\cite{FOCUS}
It is expected the error in $y$ equals to the fractional error in the lifetime
due to an excellent proper time resolution and a sensitivities of 
$1\sigma(y)=1.5\%$ is expected which is equivalent to an error 
on $r_{\rm mix}$ of $0.005\%$.

\section{Rare Decay Searches}
Charm rare decay searches are studied in flavor changing neutral current (FCNC),
lepton flavor number violation (LFNV) and lepton number violation (LNV) decays.
The FCNC decays are highly suppressed by the Glashow, 
Ilopoulos, and Maiani (GIM) mechanism and branching ratios are expected to
be $10^{-19}$-$10^{-9}$.
The LFNV and LNV are forbidden in the Standard Model.
Any results beyond the SM predictions would be a clear evidence of new physics.
Dilepton modes have experimental advantages due to clean and 
efficient lepton identification
and small combinatoric backgrounds.
To avoid any bias in selection criteria due to presence or absence of signal candidates,
all events within a signal mass window are masked.

The E791 has recently published results on searches for 24 different decay modes in
FCNC, LFNV and LNV modes.~\cite{E791Rare}
They did not observe any signals and set $90\%$ CL upper limit.
\begin{figure}[t]
\centerline{\psfig{figure=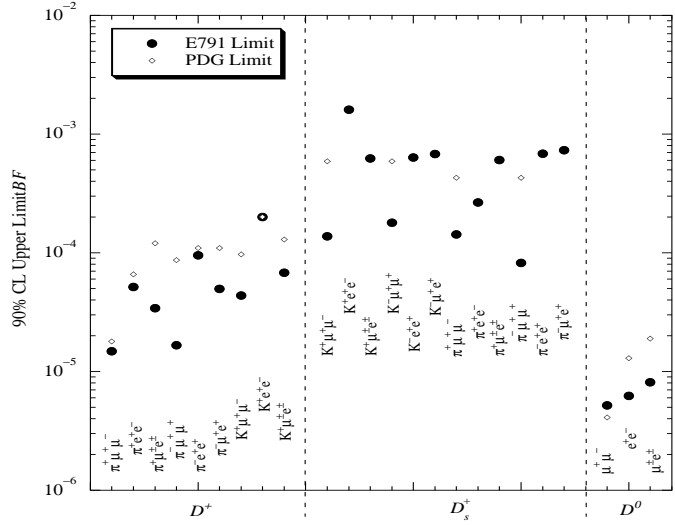,height=2.75in,width=3.5in}}
\caption{Comparision of branching fraction measurements at the
$90\%$ CL limit from E791 (dark circles) with previous best limits (open diamonds).
\label{fig:rare}}
\end{figure}
The results of searches are compared to previous limit in Fig.~\ref{fig:rare}.
8 are new measurements in $D^+_s$ and 14 are of significant improvements.
\noindent

The FOCUS is now in process of searching for forbidden and rare dilepton modes.
The blind analysis is being used in the seaches and is very similar to that
of the E791.
The current sensitivites for $D^+$ modes are at $(4$-$6)\times 10^{-6}$ level at
$90\%$ CL on upper limit.

\section{Summary}
In charm mixing search the best limits in the most general case comes 
from the CLEO measurement:
$r_{\rm mix} < 0.05\%$ at $90\%$ CL upper limit.
The E791 also has the best published limit using the semileptonic decays:
$r_{\rm mix} < 0.5\%$ at $90\%$ CL upper limit.
Analysis in progress by the FOCUS are sensitive to $r_{\rm mix}\sim 5\times 10^{-4}$.
First results from searches for a lifetime difference $\Delta\Gamma$ have
recently been published by the E791 and they find
$-0.04<y<0.06$ at $90\%$ CL.
Work in the CLEO and the FOCUS is in progress and expected sensitivies on $y$
to the few $\times 10^{-3}$ level in the very near future.
The E791 has recently published limits on the branching ratios for 24 forbidden
and rare decay modes.
In general their results are new or are of significant improvements 
over previous limits.
The rare decay searches of the FOCUS is at the early stage and their 
sensitivity to $D^+$ branching ratios is expected to be a few $\times 10^{-6}$.

\section{Acknowledgments}
Support for this work was provided by the Korea Science and Engineering Foundation.
I would like to thank George Hou and Hai-Yang Cheng for organizing an
interesting and productive meeting. Thanks also go to those members of the
ALEPH, CLEO, E791 and FOCUS collaborations who provided me with
information and plots. 
I thank Paul Sheldon, David Asner and JooSang Kang.
for many useful discussions.


\end{document}